\documentclass{article}
\usepackage{amsmath}

\newtheorem{protocol}{Protocol}

\title{A system for the simulation of simultaneous moves between two noncolocational players}
\author{Marisa Debowsky
\\
Courant Institute of Mathematical Sciences
\\
New York University
\\
251 Mercer St
\\
New York, New York  10038
\\
\\
Adrian Riskin
\\
Department of Mathematics
\\
Mary Baldwin College
\\
Staunton, Virginia  24401
\\
ariskin@mbc.edu \footnote{corresponding author}}

\begin{document}
\maketitle
\abstract{We describe a new system for the simulation of simultaneous moves
between noncolocational players.  This has applications in the burgeoning
Rock-Paper-Scissors by mail movement.}
\section{Introduction}

Intensive study of competitive Rock-Paper-Scissors (RPS) [1] has revealed a crucial psychological aspect to the game as played by colocational participants.  Expert players can spot their opponents' tells, discern predictable strategic patterns, intuit their opponents' connotational conceptions of the moves and thereby gain an edge, and so on.  In these aspects, colocational RPS (CRPS) is much like poker, and, as in the game of poker, there are many potentially sound competitors who are unable 
to play at a high level due to shyness or other psychological conditions.  Chess is another strategic game with a psychological component much like CRPS, and colocational play suffers from many of the same problems (or advantages, depending on one's point of view).  These problems are alleviated somewhat by the advent of online play, but to our minds the presence of an electronic intermediary spoils the primal purity of the games.  Although this problem seems insoluble vis a vis 
poker at this time, it is elegantly and efficiently solved in the case of chess through by-mail play.  With this situation in mind we set ourselves the task of designing an unbreakable system for by-mail RPS play that involves neither human nor electronic third-party mediation.  We have designed such a system for two players, and we describe it in the next section.  It is suitable for any game which requires simultaneous moves on the part of two players, so that it works for RPS-25 
as well as standard RPS [2]. Finally we describe a possible ranking system for federation play, such as those used in various chess federations.  We close with an invitation to the reader to join our newly founded International Postal Rock-Paper-Scissors Federation (IPRPSF).

\section{The two player system}

The game of CRPS works like this:  Alice and Bob say together ``One, two, three, shoot!''  One beat after saying the word ``shoot!" they each make one of three signs with their hands\footnote{There is a variant version where the signs are made at the same time that the word ``shoot!" is uttered, but we hold this to be mere pointless schismaticism.}: a fist (rock), a flat hand with fingers and thumb extended and held together (paper), or index and middle fingers extended and slightly apart as in the ``V-for-victory'' sign (scissors).  If the two signs thrown are the same, the round is declared a draw; otherwise, the winner is determined by the cyclic ranking: rock breaks scissors, scissors cut paper, paper covers rock.

For by-mail (non-colocational) play, Alice and Bob must communicate their moves 
to one another while each being certain that the other has made a move without knowledge of 
the opponent's move.  We reject, on aesthetic grounds, any scheme requiring online communication.  
One possible by-mail method would be for Alice and Bob to choose a particular date on which the move 
must be mailed, with the postmark verifying simultaneity.  We reject this scheme for a number of 
reasons.  First, it's too easy for things to go awry.  A letter dropped in a mailbox or picked up
 by a letter-carrier may not end up postmarked on the intended day for reasons out of the control of Alice and Bob.  It is possible to solve this problem by having the stamps hand-cancelled, but in a multi-round game this may involve a prohibitive amount of time (what with trips to the post office, waiting in line, etc.).  Furthermore, this scheme does not insure against the bribery of a postal official, as unlikely an event as that may seem.  However, the most important reason we reject this scheme is social.  One of the charms of by-mail chess is that the moves are contained in alternating letters, allowing the players to engage in a conversational correspondence as they play.  We wanted to preserve this property in a NCRPS scheme.  Finally, the scheme we came up with is much cooler than any of these other options.

The crucial observation is that with alternating rather than simultaneous play, only one player's move must be hidden.  If Alice commits to her move but does not disclose it, then Bob is free to make his move in the clear, after which point Alice's move can be revealed.  Each player has thrown a sign without knowledge of the other's move.  Thus, we simply need a mechanism by which Alice can play without revealing her move to Bob such that it can be disclosed later, ensuring that Bob can't peek before making his move and Alice can't cheat and change her move after seeing Bob's.

Our NCRPS scheme, therefore, requires a by-mail adaptation of a commitment scheme [3].  We simulate the hiding property of a commitment scheme using a sealed envelope, and we simulate the binding property using signatures across the envelope flaps.  Consider the following three-flow protocol: suppose Alice sends Bob a sealed envelope, signed across the flap.  Bob receives the envelope, does not open it, and signs below Alice's signature before sending it back to her.  Alice sees her own signature, so she knows that Bob has not replaced her original envelope with a fake one, thus ensuring that the contents of the envelope are still unknown to Bob (hiding).  If she now mails it back to Bob, he will see his own signature and know that Alice has not replaced her original envelope with a fake one, thus ensuring that she has not changed the contents (binding).  

Combining the two elements above, we present our scheme.

\begin{protocol}Alice writes her move on a piece of paper, seals it in an envelope, signs across the flap, and mails the envelope inside another envelope to Bob.  Bob signs across the flap of the envelope containing Alice's move.  Now Bob knows that as long as the envelope remains sealed Alice cannot change her her move in response to his.  Bob then returns the still-sealed envelope back to Alice with his move in the clear.  Note that at this point Alice knows who has won the game.  When she receives the envelope with the two signatures, she knows that Bob made his move without knowledge of hers.  She then returns the sealed, signed envelope to Bob, who now knows that Alice has not changed her move in response to his.  Bob now opens the sealed envelope and finds out who won the game.\end{protocol}

Traditionally, a match consists of a number of games played sequentially.  While there is a certain amount of efficiency and no mathematically identifiable advantage in Alice enclosing her next move in her last mailing of the previous game, it is possible and even likely that there is a psychological advantage.  Thus we choose to have Bob wait until Alice finishes off the first game by returning the doubly signed, still sealed envelope to him before sending her the first move of the second game, and so forth.

\section{A ranking system for Federation play}

A match consists of ten games.  Each player calculates the number of points a match is worth to him by subtracting the other player's score from his score.  For instance, if Alice wins seven games out of ten, then she gets four points on the match whereas Bob gets negative four points.  After an official match, a player's ranking is calculated using the formula shown below, where $R_{n}$ is the player's new ranking, $R_{o}$ is the player's old ranking, $P$ is the number of points the match was worth to the player, and $R_{o}^{A}$ and $R_{o}^{B}$ are Alice's and Bob's old rankings respectively.

$$R_{n} = \left\{
\begin{matrix}
{R_{o} + \left ( \left\lfloor \frac{|R_{o}^{A}-R_{o}^{B}|}{5}\right\rfloor+\frac{1}{2}\right)(P+2)} & 
{\mathrm{If \enspace lower \enspace ranked \enspace player \enspace wins \enspace or \enspace draws}}\\
 & \\
{R_{o}+\frac{1}{2}P} & {\mathrm{If \enspace higher \enspace ranked \enspace player \enspace wins}} \end{matrix}\right.$$

\noindent While this formula has not been tested extensively in federation play, it at least has the desirable property of rewarding or penalizing players more for upsets.  Each new player, by the way, is awarded an initial score of 100.

In conclusion, we would like to invite the interested reader, to whom the subtle pleasures of such an enterprise seem attractive, to submit for further information on joining the International Postal Rock Paper Scissors Federation an SASE to the corresponding author.

\section{Bibliography}
\begin{enumerate}
\item http://www.worldrps.com/
\item http://www.umop.com/rps.htm
\item Giles Brassard, David Chaum, and Claude Crepeau, {\em Minimum Disclosure Proofs of Knowledge}. Journal of Computer and System Sciences, vol. 37, pp. 156-189, 1988.
\end{enumerate}

\end{document}